\documentclass[11pt,A4paper]{amsart}
\usepackage{graphicx} 
\usepackage{amsaddr}
\usepackage{natbib}
\setcitestyle{aysep={}} 
\usepackage{hyperref} 
\usepackage{mathrsfs} 
\usepackage{tikz}
\usetikzlibrary{decorations.pathmorphing,arrows}

\makeatletter
\let\uppercasenonmath\@gobble
\makeatother

\newcommand{\tp}{\:\tilde{+}\:}
\newcommand{\RR}{\mathbb{R}}
\newcommand{\ZZ}{\mathbb{Z}}

\title{Disregarding the `Hole Argument'}
\author[Bryan W. Roberts]{Bryan W. Roberts\\\today}
\address{Philosophy, Logic \& Scientific Method\\Centre for Philosophy of Natural and Social Sciences\\London School of Economics \& Political Science\\Houghton Street, London WC2A 2AE, UK\\\href{mailto:b.w.roberts@lse.ac.uk}{b.w.roberts@lse.ac.uk}}
\email{b.w.roberts@lse.ac.uk}

\begin{document}
\begin{abstract}
  Jim Weatherall has argued that Einstein's hole argument is based on a misleading use of mathematics. I argue on the contrary that Weatherall demands an implausible restriction on how mathematics is used. The hole argument, on the other hand, is in no new danger at all.
\end{abstract}
\maketitle

\section{Introduction} 

Jim \citet{weatherall_holeargument} has argued that interpreters of general relativity may disregard Einstein's `hole argument', at least as it is presented by \citet{earmannorton1987}. He begins with the following innocuous observation: when we say two descriptions of the world are equivalent, we generally mean that they share some relevant structure. Moreover, the relevant structure for many mathematical descriptions is natural and obvious, and gets expressed as a canonical equivalence relation. This kind of thinking is familiar from a certain category-theoretic perspective, but since it may be unfamiliar for the newcomer to this perspective, let me briefly summarise this way of thinking.

Recall that the concepts of set theory are defined using just one special relation, the membership relation $\epsilon$. If two sets have the same membership relations, then this expresses a sense in which they are isomorphic. The isomorphism is established by exhibiting a mapping between two sets that preserves exactly the membership relations\footnote{The morphisms of the category of sets are just the functions, since if $f:A\rightarrow B$ is a function of sets, then $a\in A$ only if $f(a)\in f(A)$. The morphisms that preserve all and only the membership relations are the bijections.}, and in particular two sets have exactly the same membership relations if and only if there is a bijection between them. For this reason, when describing situations in which all and only set membership relations matter, it is common to say two descriptions are equivalent if and only if the sets are related by a bijection.

A similar perspective is available for Lorentzian manifolds $(M,g_{ab})$, which characterise typical models of general relativity. The facts about Lorentzian manifolds are expressed using a smooth manifold $M$ and a Lorentz-signature metric $g_{ab}$. Since these include both smoothness and metrical relations, the natural choice of morphism is a map that preserves exactly these relations; such maps are called \emph{isometries}. So, when we describe situations in which all and only the smoothness and metrical relations matter, isometries provide a natural standard of equivalence. Weatherall observes that when considering whether two descriptions are equivalent, it is important to identify the structures with respect to which equivalence is defined.

I agree. But Weatherall goes on to assert that in any legitimate mathematical representation, one must presume that mathematical structures represent a given physical scenario ``equally well, for all purposes'' if and only if they are isomorphic, where the standard of isomorphism is the one that is ``given'' by the standard formulation of a theory. I would like to point out that if Weatherall's dictum were strictly enforced, then it would be a crippling limitation on mathematical modelling. It would prohibit the human activity of freely choosing how mathematical language represents the world. And it would mean that the only legitimate way to have a good mathematical representation of is to have one that is an absolutely complete description of the world. On the contrary, even our best mathematical models in physics are invariably incomplete, and so typically fail to distinguish between distinct physical situations. Thus, for even our best mathematical models, Weatherall's dictum can dramatically fail.

\section{Regarding the Hole Argument}\label{sec:background}

\subsection{Review of Earman and Norton} A Lorentzian manifold is a pair $(M,g_{ab})$, where $M$ is a smooth manifold and $g_{ab}$ is a Lorentz-signature metric. The natural symmetries in this category are the isometries, which preserve the structure of both $M$ and $g_{ab}$: we say that $(M,g_{ab})$ and $(\tilde{M},\tilde{g}_{ab})$ are \emph{isometric} (and $\psi$ is an \emph{isometry}) if and only if there exists a diffeomorphism $\psi:M\rightarrow\tilde{M}$ such that the pushforward of $\psi$ preserves metrical relations, in that $\psi^*g_{ab}=\tilde{g}_{ab}$.

In general relativity we use Lorentzian manifolds to represent certain aspects of space and time. Often that representation involves interpreting some part of that manifold as representing a region of physical spacetime, such as a set of points containing the orbit of Mercury. But there is a further interpretive view that was advocated by Newton, called \emph{substantivalism}, which holds that spacetime refers to a substantial entity that exists independently of the matter it contains. How might one understand this statement in the language of general relativity? Earman and Norton formulate it as the view that ``the manifold is identified as spacetime and it is argued that we should hold a realist view of it'', identifying Michael \citet{friedman1983spacetime} as an endorser of the view\footnote{This perspective on substantivalism was also articulated in papers by \citet{earman1970a}, \citet{earmanfriedman1973} and \citet{sklar-spacetime}.} \citep[p.519]{earmannorton1987}. In particular they propose to evaluate any formulation of substantivalism that would deny that two isometric Lorentzian manifolds $(M,g_{ab})$ and $(\tilde{M},\tilde{g}_{ab})$ always represent the same physical situation. They called this latter statement \emph{Leibniz Equivalence}; thus, the manifold substantivalism at issue for Earman and Norton's hole argument is characterised by the denial of Leibniz Equivalence.

Manifold substantivalism is of historical interest because Einstein himself grappled with a version of it as he was coming to formulate general relativity, inventing what Earman and Norton called the Hole Argument. To highlight its modern philosophical significance, Earman and Norton formulate the argument as follows: let $(M,g_{ab})$ be a Lorentzian manifold with a Cauchy surface\footnote{A \emph{Cauchy surface} for a Lorentzian manifold $(M,g_{ab})$ is a spacelike hypersurface that is intersected by every inextendible timelike or null curve exactly once. Not all Lorentzian manifolds admit a Cauchy surface, but we can always restrict attention to a subregion $N\subseteq M$ such that $(N,g_{ab}|_{N)})$ is a Lorentzian manifold that does admit a Cauchy surface.} $S$ and a time orientation. Consider an open region $U$ to the timelike future of $S$, sometimes called the \emph{hole region}. Finally, let $\psi:M\rightarrow M$ be any diffeomorphism that is not the identity inside $U$, but which is the identity everywhere else, in such a way that the Leibniz equivalence-denier will assert that $(M,g_{ab})$ and $(M,\psi^*g_{ab})$ represent two physically distinct situations (Figure \ref{fig:hole}). If one takes $(M,g_{ab})$ and $(M,\psi^*g_{ab})$ to represent the same physical situation on the surface $S$, but different physical situations to the future of $S$, then this would be an example of indeterminism in general relativity. Thus, Earman and Norton argued, by denying Leibniz equivalence one pays the high price of introducing radical indeterminism\footnote{\citet{butterfield1989hole} and \citet{brighouse1994a} have argued that the indeterminism introduced in this way may not be so ``radical'' after all.}.

\begin{figure}[tbh]
    \caption{}\label{fig:hole}
\begin{center}
\vspace{1pc}
\begin{tikzpicture}
\draw (-1,0) -- (-3,2) node[above]{$(M,g_{ab})$} -- (-5,0);
\draw (-5,0) .. controls (-4,0.2) and (-2,-0.3) .. node[below]{{\scriptsize\;\;\;\;$S$}} (-1,0);
\draw[dotted] (-3,0.8) circle (0.6) node[right]{{\scriptsize $U$}
};
\draw[-latex] (-3,-0.1) -- (-3,1.9); 
\draw[-latex] (-3.1,-0.1) -- (-3.1,1.7); 
\draw (1,0) -- (3,2) node[above]{$(M,\psi^*g_{ab})$} -- (5,0);
\draw (5,0) .. controls (4,-0.3) and (2,0.2) .. node[below]{\scriptsize{\;\;\;\;$\psi(S)$}} (1,0); 
\draw[dotted] (3,0.8) circle (0.6) node[right]{{\scriptsize $\psi(U)$}};
\draw[-latex] (3,-0.1) -- (3,0.3) .. controls (2.8,0.8) .. (3,1.3) -- (3,1.9); 
\draw[-latex] (2.9,-0.1) -- (2.9,0.3) .. controls (2.7,0.8) .. (2.9,1.3) -- (2.9,1.7); 
\draw[-stealth'] (-1.3,1) -- node[midway,above]{$\psi$} (1.3,1);
\end{tikzpicture}
\end{center}
\end{figure}

\subsection{Disregarding the hole argument}

Weatherall's response to the hole argument hinges on a declaration about how Lorentzian manifolds represent the physical world. If two Lorentzian manifolds $(M,g_{ab})$ and $(M,\tilde{g}_{ab})$ are related by an isometry $\psi$, then according to Weatherall, %
\begin{quote}
  ``for any region $R$ of spacetime that may be adequately represented by some region $U$ of $(M,g_{ab})$, there is a corresponding region $\tilde{U} = \psi[U]$ of $(M,\tilde{g}_{ab})$ that can represent the same region of spacetime equally well, \emph{for all purposes}'' \citep[p.11][emphasis added]{weatherall_holeargument}. 
\end{quote}
Compare Weatherall's expression to that of Earman and Norton:
\begin{quote}
``\emph{Leibniz Equivalence} - Diffeomorphic models [i.e. isometric spacetimes] represent the same physical situation'' \citep[p.520]{earmannorton1987}.
\end{quote}
These two statements are very nearly the same. An important exception is Weatherall's additional locution that isometric Lorentzian manifolds represent spacetime ``equally well, for all purposes''. The only other distinction that I can make out is his use of the modal ``can'', and his consideration of an open set $U$ representing a region $R$, neither of which depart from Earman and Norton's statement in any interesting way.

However, whereas Earman and Norton take Leibniz equivalence to be an interpretive postulate to be critically evaluated, Weatherall takes his dictum to be an a priori requirement. In particular, he says that whether two mathematical models represent the same physical situation is ``given by the mathematics used in formulating those models'' (p.3), and in the particular case of Lorentzian manifolds, ``the fact that such an isometry exists provides the \emph{only} sense in which the two spacetimes are empirically equivalent'' (p.12)\footnote{And again, in his discussion of Newtonian spacetimes: ``the relevant standard of equivalence is already manifest in the map that relates the structures in the first place'' (p.18).}. To suggest otherwise would amount to ``asserting that spacetime is not or cannot be adequately represented by (precisely) a Lorentzian manifold, at least for some purposes'' (p.20), which Weatherall says cannot be right. So, by suggesting an interpretation of general relativity that denies this notion of equivalence, manifold substantivalism is dead in the water before it sets sail.

Of course, Earman and Norton also reject manifold substantivalism: this is just the hypothesis that is reduced to absurdity by the hole argument! But Weatherall denies that it is even a legitimate \emph{reductio} hypothesis to begin with\footnote{Although Weatherall later claims his view ``does not rule out (or even rule on) substantivalism or relationism'' (p.20), the precise meaning of substantivalism that Earman and Norton consider is the rejection of Leibniz equivalence, which evidently is ruled out by Weatherall's remarks, and so manifold substantivalism must be too.}, instead pillorying it as a ``misleading use of the formalism of general relativity'' (p.1), as a failure to use the formalism ``correctly, consistently, and according to our best understanding of the mathematics'' (p.3), that it is not ``mathematically natural or philosophically satisfying'' (p.21). On Weatherall's view, the hole is blocked because its reductio hypothesis is illegitimate.

Another way to understand these expressions of distaste is in terms of a kind of quietism: manifold substantivalism is so breathtakingly pointless that it is not even worth the time and energy it takes to utter. I feel the same way a few days out of the week, and will discuss quietism in the final section of this paper. But first, let me give a more critical take on the argument that Weatherall has given. 

\section{Exceptions to the given}\label{sec:critique}

\subsection{Reservations about Weatherall's argument} Weatherall has stated an interesting view, but advertises that it renders the hole argument ``blocked'' while remaining ``essentially neutral on the metaphysics of space and time'' (p.2). I am not convinced. Let me begin by discussing some reservations about the general perspective on applied mathematics that underpins his view. 

Weatherall claims that his perspective on isometric Lorentzian manifold follows from a general perspective on how mathematics can represent the world. In particular, Weatherall says he only assumes that ``isomorphic mathematical models in physics should be taken to have the same representational capacities'' in the sense that,
\begin{quote}
``if a particular mathematical model may be used to represent a physical situation, then any isomorphic model may be used to represent that situation equally well'' (p.4).
\end{quote}
This expression seems innocuous: of course one \emph{can} use two isomorphic mathematical models to represent the same physical situation. But if we interpret this expresssion strictly, then Weatherall has not in any way blocked the hole argument. Weatherall's argument requires the additional assumption that one \emph{cannot do otherwise}. In particular, he says his dictum prohibits isometric Lorentzian manifolds from representing distinct physical situations for any purpose whatsoever, not even to express manifold substantivalism for the purposes of \emph{reductio}. It is this presumption, that isometric Lorentzian manifolds represent the same situation equally well ``for all purposes'', which is supposed to prohibit the (\emph{reductio}) hypothesis that two isometric spacetimes can ever represent different things. The violation of this prohibition is supposed to render the hole argument misleading, unnatural and unsatisfying. Weatherall's argument does not go through without it. So, since Weatherall takes his critique to ``follow'' from the statement above, one can only charitably assume that he implicitly means to say ``for all purposes'' here too.

This presumption is not metaphysically neutral. On the contrary, it is a severe restriction on what mathematics can represent. Moreover, counterexamples apparently abound. Any two single-element sets are isomorphic as sets, in that there is a bijection between them. But this does not prevent one from using one single-element set to represent a black raven, while using another to represent a white shoe. Of course one \emph{can} use each of these two sets to represent the same thing (say, the raven). But nothing \emph{a priori} prevents one from using them to represent different things, and for some purposes we may wish to. Another example, co-opted from van Fraassen: two drawings of a dragon that are isomorphic as shapes in the Euclidean plane \emph{can} represent the same thing. But for some purposes (as van Fraassen suggests), we might wish to take one to represent Margaret Thatcher as draconian, while taking the other to represent a mythical beast.

The same applies to isometric Lorentzian manifolds. Of course they \emph{can} represent the same physical situation. But there is no reason to think that they must do so ``for all purposes''. In Einstein's hole argument, one considers what would happen if isometric Lorentzian manifolds were to represent different things, for the purpose of showing that a certain expression of substantivalism is absurd. Nothing \emph{a priori} prevents us from doing so, and Weatherall has given us no reason to believe otherwise.

I anticipate two sorts of rejoinders to my criticism. The first rejoinder is to say that, when we chose to represent a black raven and a white shoe in terms of single-element sets, we actually did identify them as the same. I see no way of understanding this that doesn't beg the question. I gave an example in which (for some purpose) I used single-element sets to represent different physical situations. What is wrong with that? If the answer is simply to insist that ``single-element sets represent the same situation'', then that is begging the question. The same applies to Weatherall's discussion of the hole argument: Einstein used isometric Lorentzian manifolds to represent different physical situations, for at least one purpose --- the purpose of discovering the implications of general relativity. What is wrong with that? If the response is just to declare that isometric Lorentzian manifolds represent the same physical situation ``for all purposes'', then that is not very informative.

The second rejoinder is to say that, since we think that in the real world a black raven is different from a white shoe, we were implicitly using a different mathematical structure, not single-element sets. Similarly, if someone really thinks that isometric structures can represent different physical situations, then those structures are not Lorentzian manifolds. But this response gives up the game of Weatherall's dictum: physical equivalence is not ``given by the mathematics'' automatically, but is rather determined by what we think about the real world, and by how we use mathematics. Once we accept this kind view, it is less compelling to state that manifold substantivalism cannot even be formulated as a reductio hypothesis. Manifold substantivalists are just those who think isometric Lorentzian manifolds can represent different physical situations. In fact, Weatherall considers a very similar line of response in Section 5, concluding that ``it is difficult to see how this could be done in a mathematically natural or philosophically satisfying way'' (p.21). But if that is where we have ended up, then Earman and Norton's hole argument was never ``blocked'' after all. After all, they conclude a very similar similar thing: manifold substantivalism is indeed unnatural and unsatisfying, in the sense that it introduces a radical form of indeterminism into general relativity.

\begin{figure}[bth]
    \caption{}\label{fig:embedding}
\begin{center}
\vspace{1pc}
\begin{tikzpicture}[scale=0.75]
  \draw[style=help lines, step=0.5] (-5,-1.99) grid (-1,2);
  \draw[decorate,decoration={snake, segment length=6pt, amplitude=2pt}] (-5,-2) -- (-1,-2);
   \node at (0,0.5) {$\psi_s$};
   \draw[line width=0.5pt,-stealth'] (-5,-2) -- (0.9,-1);
   \draw[line width=0.5pt,-stealth'] (-5,1) -- (0.9,2);
  \draw[style=help lines, step=0.5] (0.99,-0.99) grid (5,2);
  \draw[dashed] (1,-1) -- (5,-1);
  \draw[style=help lines] (1,-1) -- (1,-2); 
  \draw[style=help lines] (5,-1) -- (5,-2);
  \draw[decorate,decoration={snake, segment length=6pt, amplitude=2pt}] (1,-2) -- (5,-2);  

\end{tikzpicture}
\end{center}
\end{figure}

Here is another example, more concretely in the context of general relativity, to suggest that we should not be so quick to say that ``the fact that such an isometry exists provides the \emph{only} sense in which the two spacetimes are empirically equivalent'' \citep[p.12]{weatherall_holeargument}. Let $(M,\eta_{ab})$ be the half-plane $M=\RR\times(0,\infty)$ together with the Minkowski metric. Then there is a semigroup of isometries $\psi_s:\RR\times(0,\infty)\rightarrow \RR\times(s,\infty)$ that embed $(M,\eta_{ab})$ into a proper subset of itself, as illustrated in Figure \ref{fig:embedding}. Weatherall's view appears to imply not only that these two Lorentzian manifolds \emph{can} represent the same physical situation, but that they do so ``equally well, for all purposes'', in spite of the fact that one is a proper part of the other. It is the ``for all purposes'' part that seems to me to be too much. It is perfectly coherent to distinguish what happens in an arbitrary region in spacetime from what happens in a proper subregion, in spite of the fact that the Lorentzian manifolds representing these regions are related by an isometry. Of course for some purposes one \emph{can} use both Lorentzian manifolds to represent the same situation. But we might also wish to treat them as representing different physical situations, at least for some purposes. It is in no way ``unnatural'' or ``unsatisfying'' to distinguish events that occur in a proper submanifold of $(M,\eta_{ab})$ from those that occur in $(M,\eta_{ab})$ more generally.

I suspect that some readers may find my drilling of the ``for all purposes'' locution confusing, when they compare this response to some passages in Weatherall like the following.
\begin{quote}
On the view described there, once one asserts that spacetime is represent by a Lorentzian manifold, one is committed to taking isometric spacetimes to have the capacity to represent the same physical situations, since isometry is the standard of isomorphism given in the mathematical theory of Lorentzian manifolds. To deny this would be in effect to insist that it is some other structure --- one that is not preserved by isometries --- that represents spacetime in relativity theory. \citep[p.20]{weatherall_holeargument}
\end{quote}
Here again the ``for all purposes'' locution has been dropped. Other than that, the passage describes options very similar to the two rejoinders I considered above. But I do not see how the argument can work without something like the addition of ``for all purposes''. The hole argument does not consider whether two isometric spacetimes merely have ``the capacity'' to represent the same physical situation. The question is whether one can say that there is \emph{some} situation in which the manifold substantivalist will deny that they are the same. Weatherall blocks the hole argument by rejecting this statement, on grounds that isometric spacetimes ``represent the same region of spacetime equally well, for all purposes" (p.11) and that "the fact that such an isometry exists provides the \emph{only} sense in which two spacetimes are empirically equivalent'' (p.12). This is the main argument, and it is this that I have argued is implausible.


\subsection{Alternative takes on representation}

It is difficult to maintain that equivalence in empirical science is somehow ``given'' by the mathematical language we use to describe it, or that some standard of isomorphism is the only way to describe empirical equivalence. Representing the world mathematically involves formulating particular human intentions about how that representation is supposed to work. We must take these considerations into account when deciding whether or not two descriptions are empirically equivalent.

The equivalence relation we use to say when two physical situations are the same in general depends on what we would like a Lorentzian manifold to represent. Isometry is not ``the only way''. For example, Lorentzian manifolds have been used to describe the totality of knowledge that one can in principle acquire. This thinking led \citet{glymour1972a,glymour1977a}, \citet{malament1977b}, and \citet{manchak2008prediction} to an alternative standard of empirical equivalence: two spacetimes $(M,g_{ab})$ and $(\tilde{M},\tilde{g}_{ab})$ are \emph{observationally equivalent} if and only if for each point $p\in M$ there exists a $\tilde{p}\in\tilde{M}$ such that the backwards light cones $I^-(p)$ and $I^-(\tilde{p})$ are related by an isometry. Of course, this means that every pair of isometric spacetimes is also observationally equivalent. But the converse is generally false: \citet{manchak2008prediction} proved that all but a few bizarre Lorentzian manifolds admit observationally equivalent counterparts to which they are not isometric. This notion of empirical equivalence, though weaker than that of an isometry, is an interesting standard of equivalence for both philosophical and physical practice. It is an equivalence relation determined by isometries between backward light cones. There is nothing wrong with this because the choice equivalence of physical equivalence relation is a matter of human convention, not a priori fact.

Another notion of equivalence arises when we take a Lorentzian manifold to represent spacetime accurately only under certain circumstances, such as when the interaction between gravitation and quantum fields can be ignored with negligible error. In other words, we may have good reason to think a representation is incomplete. Wigner believed that such incompleteness is generic among laws of physics, writing that ``laws of nature contain, in even their remotest consequences, only a small part of our knowledge of the inanimate world'' \cite[p.5]{wigner1960unreasonable}. But this kind of incompleteness does not prevent us from legitimately describing spacetime using a Lorentzian manifold.

\begin{figure}[tbh]
    \caption{}\label{fig:blackholes}
\begin{center}
\vspace{1pc}
\begin{tikzpicture}[scale=0.4]
  \draw (0,1) -- (0,-8) -- node[midway,right] {\scriptsize $\mathscr{I}^-$} (8,0) --  node[midway,right] {\scriptsize $\mathscr{I}^+$} (4,4) -- (4,1);
  \draw[decorate,decoration={snake, segment length=6pt, amplitude=2pt}] (0,1) -- (4,1);
  \draw[densely dotted] (4,1) -- (0,-3);
\end{tikzpicture}
\end{center}
\end{figure}
Such as in the case with the Lorentzian manifold representing black hole evaporation, pictured in Figure \ref{fig:blackholes} as a conformal diagram. As a Lorentzian manifold $(M,g_{ab})$, this structure is almost universally understood to be incomplete, and many competing accounts of what it represents have been given. For example, many take the evolution of a quantum field over the course of the black hole's history to evolve unitarily according to the laws of quantum theory, while others take that evolution to be non-unitary. This practice proceeds with the understanding that isometry (and even identity!) is not ``the only way'' to say whether two physical spacetimes are equivalent, since Lorentzian manifolds represent spacetime incompletely. But this should not dissuade anyone from the practice of representing spacetime using a Lorentzian manifold. We must simply admit that two isometric spacetimes --- in this case, the very same spacetime --- may sometimes be used to represent different physical situations.

Even the hole argument itself was once used for the purposes of developing new physics in this way. Einstein himself discovered a version of the argument while trying to formulate general relativity. There is no reason to prohibit Einstein's thinking, or to prohibit the more general practice of exploring when two mathematical structures represent the same situation, whether or not those structures are isomorphic. To do so would be overly restrictive, and perhaps even skirting the dreaded practice of \emph{a priori} physics.

\section{On warming up}\label{sec:warmup}

Weatherall argues that his perspective on the hole argument can be motivated by a warm-up exercise. This exercise is susceptible to the same problems as Weatherall's main argument.

\subsection{Weatherall's integer exercise} In the exercise we consider two more abstract structures, the groups of integers, $(\ZZ,+)$ and $(\ZZ,\tilde{+})$. The binary operation `$+$' of the first group is normal arithmetic addition, so that $3+5=8$, etc. The binary operation `$\tilde{+}$' of the second group is arithmetic addition followed by subtraction of 1, so that $3 \tp 5=7$, and in general $n \tp m = n+m-1$.

Weatherall asks whether there is an ambiguity with regard to which number is the identity in the group of integers. The identity element of the first group $(\ZZ,+)$ is $0$, since $0+n=n+0=0$ for all $n\in\ZZ$. The identity of the second group $(\ZZ,\tilde{+})$ is $1$, since $1 \tp n = n \tp 1 = n$ for all $n\in\ZZ$. He concludes, quite correctly, that there is no ambiguity when `the identity' is interpreted as either the group identity or the set element $0$. If `identity' means `group identity,' then the term is only meaningful relative to a chosen group. So, $0$ is the identity for $(\ZZ,+)$ and $1$ is the identity for $(\ZZ,\tilde{+})$. On the other hand, if one means `the set element $n$' where (say) $n=0$, then again there is no ambiguity in specifying this object. Thus, each concept of identity conferred by the formalism provides an unambiguous way to represent the numbers.

All this is perfectly fine. But set theory and group theory are not the only tools available for distinguishing numbers. One particularly relevant alternative arises when one presumes a certain kind of realism about numbers known as a \emph{platonism}. For the platonist, a number like (say) zero may have a mind-independent existence, which provides the definitive criterion for whether or not a given object $n$ can serve as its representative. In other words, the platonist uses mathematical reality, not group or set structure, in order to determine whether two representations are equivalent.

I do not wish to advocate mathematical platonism. The point I am making is that \emph{if} one is a realist about some objects, then that realism may provide an independent sense in which two such objects are or or not the same. Just as I may take the letter $S$ to denote the set of planets in the universe that have no moons, so the realist about numbers can take the set $\ZZ$ to denote the set objects in the mathematical universe that are integers. For this kind of realist, there is a meaningful question about whether the identity of the group $(\ZZ,+)$ or the identity of the group $(\ZZ,\tilde{+})$ corresponds to a given number $0$ in the mind-independent set $\ZZ$. Those of us that are not platonists will find this question strange, but the point is that it is a meaningful question to the realist, and it is the kind of question that comes up when we use language to describe the real world. The fact that there is no ambiguity about the group identity of each is irrelevant.

Let me point out an alternative argument using Weatherall's example, which I think better sets the stage for the hole argument\footnote{This argument is akin to a classic argument of \citet{benacerraf1965sets}, and related ones discussed by \cite[Ch.6]{kitcher1984nature} and \citet[Ch.10]{shapiro2000thinking}.}. It is an argument against the plausibility of realism about the integers, which runs as follows.

\begin{enumerate}
	\item Suppose for reductio that the integers $\ZZ$ have a mind-independent existence, and as a matter of mind-independent fact form a group isomorphic to $(\ZZ,+)$, and with additive identity $0$.
	\item Define the groups of integers $(\ZZ,+)$ and $(\ZZ,\tilde{+})$ as above, with additive identities $0$ and $1$, respectively.
	\item Observe that the group theoretic structure of these groups alone does not determine which (if either) of $0$ or $1$ is the true identity. Thus, realism about integers violates the principle (which someone might call `Group Equivalence'\footnote{As the name suggests, Group Equivalence is akin to Leibniz Equivalence in the discussion of the Hole Argument.}) that every group isomorphism relates equivalent mathematical states of affairs.
	\item This failure may be too high a price to pay for a metaphysical view about numbers. Thus, realism about integers is implausible.
\end{enumerate}

This argument is a much closer analogue of the hole argument. Of course, the final step here is questionable; I do not think the failure of `Group Equivalence' is a convincing refutation of realism about numbers. But the analogous failure in the hole argument is much more dramatic, corresponding to a radical failure of a certain kind of Laplacian determinism\footnote{\citet{butterfield1989hole} exhibits an interesting sense of determinism that still be saved.}.

\subsection{Rotations of a vector}\label{subs:rotations}

An alternative perspective can be gained with some further warming up. Let me propose a similar exercise that makes use of a more concrete structure, the rotations of the vector $v$ shown in Figure \ref{fig:vectors}.

\begin{figure}[hbt]
\begin{center}
\begin{tikzpicture}[scale=0.75]
  \draw[style=help lines, step=0.5] (-1.9,-1.9) grid (1.9,1.9);
  \draw[line width=1.5pt,-stealth'] (-1.5,-1.5) node[circle,fill,inner sep=1.5pt]{} -- (2,-1.5) node[label=right:$v$]{};
\end{tikzpicture}
$\;\;$
\begin{tikzpicture}[scale=0.75]
  \draw[style=help lines, step=0.5] (-1.9,-1.9) grid (1.9,1.9);
  \draw[line width=1.5pt,-stealth'] (-1.5,-1.5) node[circle,fill,inner sep=1.5pt]{} -- (1.53,0.25) node[label=right:$R_{\theta}v$]{};
  \draw [thick,domain=0:25,->] plot ({-1.5+3*cos(\x)}, {-1.5+3*sin(\x)});
  \draw [thick,domain=0:25,->] plot ({-1.5+2*cos(\x)}, {-1.5+2*sin(\x)});
\end{tikzpicture}
\end{center}
\caption{}\label{fig:vectors}
\end{figure}

We understand what it means to rotate this figure. No mathematics is needed for that; we can just pick up the page and turn it. Call these the physical rotations. These physical rotations may be expressed in mathematical language. For example, writing $v\in\RR^2$ in Cartesian coordinates, we can define a group of rotations using the set of matrices $\{R_\theta : \theta \in[0,2\pi)\}$ under the operation of matrix multiplication, where,
\[
	R_\theta = \left( \begin{array}{cc}
		\cos\theta & -\sin\theta \\
		\sin\theta & \cos\theta
	\end{array} \right).
\]
Each matrix $R_\theta$ transforms a vector $v$ to one that is rotated counterclockwise through the angle $\theta$. These matrices have the properties that $R_\theta R_{\theta^\prime}=R_{\theta+\theta^\prime}$, and that the identity matrix is $R_0=I$, in that $IR_\theta=R_\theta I = R_\theta$ for all $R_\theta$. The structure $(R_\theta, \cdot)$ forms a group, which provides one precise mathematical representation of the physical rotations of the arrow.

However, there are many ways to instantiate a rotation group. Let me define a new binary operation `$\ast$' on the same matrices by the relation
\[
  R_\theta\ast R_{\theta^\prime}:=R_{\theta+\theta^\prime-\pi}.
\]
The identity element for the new group $(R_\theta,\ast)$ is not the identity matrix, but rather $R_{\pi}$, since $R_{\pi}\ast R_\theta = R_\theta\ast R_\pi = R_\theta$ for all rotations $R_\theta$.

There is an isomorphism from $(R_\theta, \ast)$ to $(R_\theta, \cdot)$ given by $\rho(R_\theta)=R_{\theta-\pi}$. We thus have two isomorphic groups defined on the same underlying set of rotation matrices. Does this imply that both are equally correct ways to describe the physical rotations of the vector $v$? Or that the matrices $I$ and $R_{\pi}$ are equally correct representatives of the identity rotation? Of course not: the first description (with the standard identity element $I$) is correct in a way that the second description is not. We can say why this is without referring to any special mathematical objects: we began with an experience of what it means to physically rotate the vector $v$, and our description in terms of the second group fails to adequately capture that experience.

Weatherall's dictum prohibits us from making any such judgement. The two descriptions of the physical rotations are given by isomorphic mathematical models, which therefore represent a given physical situation ``equally well, for all purposes''. Thus one must seemingly conclude that both descriptions are equally good models of the physical rotations. This is a strange conclusion to arrive at. When representing the world in terms of groups, as with many mathematical structures, there may be reasons external to the formalism that lead us to distinguish between isomorphic models. To prohibit any such distinction would be an implausible restriction on how mathematics is used.

One might be tempted to try to save Weatherall's dictum by adding more mathematical structure. For example, instead of describing the rotations of the arrow using the group $G=(R_\theta,\cdot)$, one could describe them using a \emph{matrix representation}, which is a pair $(G,\rho)$ with $\rho:G\rightarrow GL$ a homomorphism from $G$ into the `General Linear' group $GL$ of $2\times 2$ matrices over the real numbers. The first group $G=(R_\theta,\cdot)$ is made into a matrix representation by introducing the mapping $\iota: R_\theta\mapsto R_\theta$. The second group $G^*=(R_\theta,\ast)$ is made into a matrix representation by introducing the mapping $\rho:R_\theta:\mapsto R_{\theta-\pi}$. As matrix representations, the structures $(G,\iota)$ and $(G^*,\rho)$ are the same. They both take the group identity to the matrix identity $I$. They both take the order-2 group element to $R_\pi$. Thus, on this more elaborate description of rotations, we have just one representation of the rotation group, and the previous difficulty does not arise.

I do not deny that a matrix representations provide one way to describe physical rotations. But this does nothing to improve the plausibility of Weatherall's dictum. Our concern here is with the common-language use of `representation' as a model or description of a physical situation. This should not be conflated with the mathematically precise concept of a matrix representation. The groups $G=(R_\theta,\cdot)$ and $G^*=(R_\theta,\ast)$ each provide a representation of the physical rotations in the common-language sense, even though neither has been given enough structure to count as a matrix representation.

This might lead one to observe that a matrix representation describes the physical rotations in a more complete way than a mere group. Problems only arise for Weatherall's dictum because a group alone provides an incomplete description of the physical rotations. But that is exactly the point. Factors outside of a given mathematical formalism may distinguish between two descriptions, even when the formalism itself does not. The point that I would like to emphasise is that this is not a rare occurrence, but a very common feature of mathematical modelling. One can avoid the trap by reminding oneself that, contrary to what Weatherall suggests, it is Nature that ultimately determines what is physically equivalent and what is not.

\section{An alternative brand of quietism}\label{sec:quietism}

There is gentler brand of quietism in the neighborhood of Weatherall's view that might be worth clarifying. It is an attitude that I myself adopt from time to time, and provides some guidance on how to react to the Hole Argument. The main difference is that this view will be presented as a mere attitude, as opposed to a rule restricting the use of mathematical representations. I know of no argument that establishes the present view. Some simply take comfort in the gentle, Buddhist-like perspective on the philosophy of physics that this attitude provides.

The attitude begins by stating the propositions that we have good evidence to believe, in the normal language of science\footnote{This is what \citet[\S 1]{ruetsche-qt} refers to as a \emph{partial} interpretation of a theory.}. For example, we may all agree that the region near the galactic centre has the structure of Kerr spacetime. But at this point, the attitude refuses all further interpretive claims. Questions like `Is the manifold $M$ is real?' are passed over silently. In their stead one adopts an attitude of quietism as far as the propositions of realism about unobservables are concerned.

I take this to capture a sense of what Arthur Fine has called the Natural Ontological Attitude (NOA), which he summarizes as the recommendation to `try to take science on its own terms, and try not to read things into science' \cite[p.149]{fine-shakygame}. This perspective can be helpful, and indeed I often find myself joining its practitioners in the monastery for a little peace of mind. However, there is no use pretending that this view is established by any rigorous argument or rule, as Fine is quick to point out:
\begin{quote}
It does not comprise a doctrine, nor does it set a philosophical agenda. At most it orients us somewhat on how to pursue problems of interest, promoting some issues relative to others just because they more clearly connect with science itself. Such a redirection is exactly what we want and expect from an attitude, which is all that NOA advertises itself as being. \cite[p.10]{fine-shakygame}.
\end{quote}
The NOA attitude toward manifold substantivalism, I take it, is an exercise in the discipline of silence.

However, the hole argument is not necessarily a case where this attitude is appropriate. The argument itself identifies an interesting connection between the realism debate and philosophy of science, in establishing a link between manifold substantivalism and indeterminism. It has also promoted interesting connections between the realism debate and modern physics in helping to motivate a relationist perspective on spacetime in quantum gravity \citep{isham1993canonical,earmanbelot1999,rovelli-quantumgravity} as well as sophisticated substantivalist alternatives \citep{pooley2006}. Clearly, with too much quietism you may miss out on some of the fun. But a healthy dose of the kind of quietism I have described here may still help orient one toward the more fruitful problems in the philosophy of physics.

\section{Conclusion}\label{sec:conclusion}

Formal equivalence relations are only meaningful once a standard of equivalence has been identified. But it would be a mistake to suggest that the only acceptable equivalence relations are those ``given'' by the formalism of a physical theory. The hole argument has not been `blocked' by Weatherall's discussion, and previous commentators have not failed to `recognize the mathematical significance of an isomorphism'\citep[p.13 fn.20]{weatherall_holeargument}. It simply concerns matters of realism that are precluded from Weatherall's discussion by fiat. The danger in such a radical restriction on mathematical representation is that it forgoes the practice of using mathematical representations that are incomplete. \cite{earmannorton1987} argued that as a metaphysical doctrine, manifold substantivalism may come at too high a price. The price of Weatherall's dictum for applied mathematics may be at least as high.

\section*{Acknowledgements}

Thanks to Jim Weatherall for several interesting and lively discussions on this topic, as well as Jeremy Butterfield, Bal\'azs Gyenis, James Nguyen, John D. Norton and Oliver Pooley for valuable comments.

\end{document}